\documentclass[prd,twocolumn,showpacs,amsmath,amssymb]{revtex4}
\usepackage{graphicx}% Include figure files
\usepackage{epstopdf}
\usepackage{bm}% bold math
\usepackage{hyperref}% makes the references hyperlinked
\begin{document}
%\preprint{{\tiny CECS-PHY-09/06} }
\title{Phase transitions in  charged topological black holes dressed with a scalar hair}
\author{Cristi\'an Mart\'{\i}nez}\email{martinez@cecs.cl}
\affiliation{Centro de Estudios Cient\'{\i}ficos (CECS),
 Casilla 1469, Valdivia, Chile\\and
Centro de Ingenier\'{\i}a de la Innovaci\'on del CECS (CIN),
Valdivia, Chile}
\author{Alejandra Montecinos}\email{alejandramontecinos@unab.cl}
\affiliation{Departamento de Ciencias F\'{\i}sicas, Universidad  Andr\'es Bello, Av. Rep\'ublica 239, Santiago, Chile}

\begin{abstract}
\noindent Phase transitions in charged topological black holes dressed with a scalar field are studied.  These black holes are solutions of the Einstein-Maxwell theory with a negative cosmological constant and a conformally coupled real self-interacting scalar field.  Comparing, in the grand canonical ensemble, the free energies of the hairy and undressed black holes two different phase transitions are found. The first of them is one of second-order type and it occurs at a temperature defined by the value of the cosmological constant. Below this temperature an undressed black hole spontaneously acquires a scalar hair.  The other phase transition is one of first-order type. The corresponding critical temperature, which is bounded from above by the one of the previous case, strongly depends on the coupling constant of the quartic self-interaction potential, and this transition only appears  when the coupling constant is less than a certain value. In this case, below the critical temperature the undressed black is thermodynamically favored. However,  when the temperature exceeds the critical value a hairy black hole is likely to be occur.
\end{abstract}
\maketitle

\section{Introduction}

Thermodynamics of black holes is one of the most interesting research topic, given that  the thermodynamical properties of these fascinating objects are based on its elusive quantum description. One can consider black holes as  states in a thermodynamical ensemble and to compare the free energy associated to each of them. In some cases phase transitions have been found. A  well-known example is due to Hawking and Page \cite{Hawking:1982dh}. Recently,  the study of phase transitions of  black holes endowed of a scalar hair has focused much interest since these transitions have been related with holographic superconductivity \cite{Gubser:2008px,Hartnoll:2008vx} in the context of the AdS/CFT correspondence  (see relevant reviews in \cite{reviews}).  In general,  the no-hair theorems rule out such a class of solutions. However, when a negative cosmological constant is considered, there exist exact black holes with a scalar hair and, consequently, phase transitions can be analytically studied. The first of these hairy black holes was obtained  in three spacetime dimensions \cite{Martinez:1996gn} and then generalized  \cite{Henneaux:2002wm} by considering an one-parameter self-interaction potential.  It was found \cite{Gegenberg:2003jr} that the scalar hairy black hole of Ref. \cite{Henneaux:2002wm} can decay into the BTZ black hole irrespective of the horizon radius. The presence of a cosmological constant also allows the existence of event horizons with non-trivial topologies. Thus, a four-dimensional \textit{topological} black hole dressed with a scalar field, whose  event horizon is a compact space of negative curvature,  was introduced in \cite{Martinez:2004nb}. Moreover, it was shown the existence of a second-order phase transition at a critical temperature, below which a black hole in vacuum undergoes a spontaneous dressing up with a nontrivial scalar field. The order parameter that characterizes this transition depends on the value of the scalar field at the horizon. These results were reproduced in \cite{Koutsoumbas:2006xj,Myung:2008ze}, and notably, it was found \cite{Koutsoumbas:2009pa} that this phase transition is dual to a superconductivity transition. The existence of this second-order phase transition seems to be more general \cite{Kolyvaris:2009pc}. 

These hairy black holes can be charged by adding a Maxwell gauge field. Thus, an electrically charged black hole with a conformally coupled scalar field was obtained in \cite{Martinez:2005di} and one with a minimally coupled scalar  in \cite{Martinez:2006an}. In the minimally coupled case, phase transitions are not possible between the hairy and the non-extremal hyperbolic  Reissner-Nordstr{\"o}m (hRN) black holes, being the latter one  thermodynamically favored. However,  there is a transition where the extremal black hole without scalar field is endowed with a non-trivial scalar hair \cite{Martinez:2006an}.  In this report we focus the analysis in the case of the charged  black hole with a conformally coupled scalar field \cite{Martinez:2005di}.   The main properties of this exact solution, hereafter named \textit{hairy black hole}, will be reviewed in the next section. In advance, we note that the coupling constant $\alpha$ of the quartic self-interaction potential plays an important role in this analysis. Because we have an exact solution,  it is interesting enough to explore the thermodynamical transitions of this hairy black hole. This is the aim of this brief report.  We apprise to the readers that the question about how these phase transitions are or not related to a superconductivity transition will be not discussed here.  

Previous attempts have dealt with to describe phase transitions in the hairy black hole. However, these treatments have some problems. In \cite{Koutsoumbas:2008pw} the free energy is computed using an expression for the entropy  with an incorrect factor. In this case the scalar field is conformally  coupled to gravity. Consequently, in the area law for the entropy an ``effective" Newton constant, depending on the value scalar field at the horizon,  must be considered. More recently, it was argued that  a second-order transition from a hairy to  a hRN black hole is unlikely possible to occur \cite{Myung:2010rb}. However, the analysis in \cite{Myung:2010rb} is not clear for lack of a thermodynamical ensemble. In order to solve these problems, we develop a detailed study of phase transitions for the hairy black hole in Section \ref{transiciones}. The results show the existence of two phase transitions of different order. One of them corresponds to a second order transition, which occurs at a fixed temperature defined only in terms of the cosmological constant. On contrary, the other transition is of first order and the critical temperature strongly depends on $\alpha$.

%%%%%%%%%%%%%%%%%%%%%%%%
\section{Charged topological  black holes dressed with a scalar field }\label{Sols}
%%%%%%%%%%%%%%%%%%%%%%%%

In this section, we briefly review the main properties of hairy and hRN black holes.  We start considering four-dimensional gravity with a cosmological constant, where the matter content is given by a conformally coupled real self-interacting scalar field and a Maxwell gauge field. The action is given by
\begin{eqnarray}
I[g_{\mu \nu},A_{\mu}]&=&\int d^4x\sqrt{-g}\left[\frac{R-2\Lambda}{16\pi G}-\frac{F^{\mu\nu}F_{\mu\nu}}{16\pi} \right.\nonumber\\
&-&\left. \frac{1}{2}g^{\mu\nu} \partial_\mu\phi\partial_\nu\phi-\frac{1}{12}R\phi^2-\alpha\phi^4\right] \label{action}
\end{eqnarray}
with $F_{\mu\nu}=\partial_{\mu} A_{\nu}-\partial_{\nu} A_{\mu}$. When the cosmological constant is negative, $\Lambda=-3/l^2$, the
theory admits an exact solution \cite{Martinez:2005di}, which we have called hairy black hole, and whose line element is  
\begin{eqnarray}\label{solu}
ds^2&=&-\left[\frac{r^2}{l^2}-\left(1+\frac{G\mu}{r}\right)^2\right]dt^2 \nonumber \\&+&\left[\frac{r^2}{l^2}-\left(1+\frac{G\mu}{r}\right)^2\right]^{-1}dr^2 +r^2d\sigma^2.
\end{eqnarray}
In the above expression $d\sigma^2$ is the line element of a negative constant curvature
two-dimensional manifold, which is assumed to be compact and with volume $\sigma$.  The scalar field  reads
\begin{equation}
\phi=\sqrt\frac{1}{2\alpha l^2}\left(\frac{G\mu}{r+G\mu}\right),
\end{equation}
and the electromagnetic potential has the simple form
\begin{equation}
A=A_0(r) dt=-\frac{q}{r}dt.
\end{equation}
The constants $q$ and $\mu$ are not independent, they are related in the following way
\begin{equation}\label{carga}
q^2=-G\mu^2\left(1-\frac{2\pi G}{3l^2\alpha}\right)\equiv G\mu^2(a-1) ,
\end{equation}
where $a=2\pi G/(3l^2\alpha) \geq1$ because $q$ is real.

If $\mu$ is positive the black hole has a single horizon. On the other hand, in the case
$-l/4\leq G\mu < 0$ the solution have three horizons. The
extreme black hole occurs for $G\mu=-l/4$, where two horizons take the
value $l/2$.

Other solution of this theory is the hRN black hole \cite{Brill:1997mf}, which is  obtained from (\ref{action}) provided that $\phi\equiv0$. The corresponding metric is  
\begin{eqnarray}\label{solu2}
ds^2=&-&\left(\frac{\rho^2}{l^2}-1-\frac{2G\mu_0}{\rho}+\frac{Gq_0^2}{\rho^2}\right) dt^2\nonumber\\
&+&\left(\frac{\rho^2}{l^2}-1-\frac{2G\mu_0}{\rho}+\frac{G
q_0^2}{\rho^2}\right)^{-1}d\rho^2+\rho^2d\sigma^2,
\end{eqnarray}
where $\rho >0$. The integration constants  $\mu_0$ and $q_0$ are independent and they are proportional to the mass and electric charge respectively.  The electromagnetic potential read $A=-(q_0/\rho) dt$. The event horizon is located at $\rho_+$, which satisfies the equation
\begin{equation}
\frac{\rho_+^2}{l^2}-1-\frac{2 G\mu_0 }{\rho_{+}}+\frac{G q_0^2 }{\rho_{+}^{2}}=0.
\end{equation}
The hRN black hole becomes an extremal one when $\mu_0$ and $q_0$ hold the following relations
\begin{equation} \label{extremo}
G\mu_0 |_{\textrm{ext}}= \left(\frac{2\rho_+^2}{l^2}-1\right) \rho_{+},\quad G q_0^2 |_{\textrm{ext}}=\left(\frac{3\rho_+^2}{l^2}-1\right)\rho_{+}^2.
\end{equation} 

%%%%%%%%%%%%%%%%%%%%%%%%%%%%%%%%%
\section{Phase transitions}\label{transiciones}
%%%%%%%%%%%%%%%%%%%%%%%%%%%%%%%%%

In this section, we will compare the free energy of hairy black and hRN black holes. In \cite{Martinez:2005di}, the free energy, as well as other thermodynamical quantities, were computed for the hairy black hole using the grand canonical ensemble. In this ensemble the temperature $T$, or equivalently its inverse $\beta$,  and the potential difference  $\Phi=A_0(\infty)-A_0(\textrm {horizon})$ are fixed. Then, the free energy $F$ is a function of $T$ and $\Phi$ and it has the following expression 
\begin{equation}\label{freegeneral}
F=F(T,\Phi)=M-T S-\Phi Q
\end{equation}
where $M$, $Q$, $S$  denote the mass,  the electric charge, and the entropy, respectively.  For the hairy black hole, we have 
\begin{equation} \label{TQcMTZ}
T=\frac{1}{2\pi l}\left(\frac{2 r_+}{l}-1\right), \quad \Phi= \frac{q}{r_+}, 
\end{equation}
and
\begin{equation}
M= \frac{\sigma}{4\pi}\mu, \quad  Q=\frac{\sigma}{4\pi}q, \quad  S=\frac{\sigma r_{+}^{2}}{4G^*},
\end{equation}
where  $G^*$ stands for the effective Newton constant
\begin{equation} 
\frac{1}{G^*}\equiv \frac{1}{G}\left(1-\frac{4\pi G}{3}\phi^2(r_+)\right).
\end{equation}
As it was noted in \cite{Martinez:2005di}, the entropy is positive provided that $\mu$ is bounded by specific values, which depend on $a$. In terms of the temperature, this  condition states that it must range in the interval $(T_{min},T_{max})$, where 
\begin{equation}
T_{min}= \frac{1}{2 \pi l}\left( \frac{\sqrt{a}-1}{\sqrt{a}+1}\right), \quad T_{max}= \frac{1}{2 \pi l}\left( \frac{\sqrt{a}+1}{\sqrt{a}-1}\right).
\end{equation}

For the hRN black hole, the thermodynamical variables are given by
\begin{equation} \label{TQhRN}
T=\frac{1}{2\pi l}\left(\frac{\rho_+}{l}+\frac{G\mu_0 l}{\rho_{+}^{2}}-\frac{Gq_0^2 l}{\rho_{+}^{3}}\right), \quad \Phi= \frac{q_0}{\rho_+}, 
\end{equation}
and
\begin{equation}
M= \frac{\sigma}{4\pi}\mu_0, \quad  Q=\frac{\sigma}{4\pi}q_0, \quad  S=\frac{\sigma \rho_{+}^{2}}{4G}.
\end{equation}
In order to find phase transitions between the hairy and undressed states, we must consider both black holes in a same grand canonical ensemble, i.e. at the same $T$ and $\Phi$. Equaling  $T$ and $\Phi$  for both black holes, from Eqs. (\ref{TQcMTZ}) and (\ref{TQhRN}) we obtain the conditions
\begin{equation} \label{equal}
\frac{2 r_+}{l}-1=\frac{\rho_+}{l}+\frac{G\mu_0 l}{\rho_{+}^{2}}-\frac{Gq_0^2 l}{\rho_{+}^{3}}, \quad \mbox{and} \quad
 \frac{q}{r_+}= \frac{q_0}{\rho_+}.
\end{equation}
The above conditions become identities when one express, using Eqs. (\ref{TQcMTZ}) and (\ref{TQhRN}),  $r_+$, $q$, $\rho_{+}$, and $q_0$ in terms of $T$ and $\Phi$. However, the algebraic manipulation is more simple if Eqs. (\ref{TQcMTZ}), (\ref{TQhRN}) and (\ref{equal}) are used at the same time. Thus, in this setup, the corresponding free energies are: 
\begin{eqnarray}\label{delta}
\!\!F_{\textrm{hRN}}&\!\!=\!\!&-\frac{\sigma
l}{8\pi G }\!\left[2\left(\frac{\rho_+}{l}-\pi l T\right)\frac{\rho_+^2}{l^2} \right]  \\
\!\!F_{\textrm{hairy}}&\!\!=\!\!&-\frac{\sigma
l}{8\pi G }\!\left[ \left(\pi l T+\frac{1}{2}\right)^2+a\left(\pi l T-\frac{1}{2}\right)^2\right]\!\!,
\end{eqnarray}
with
\begin{equation}\label{rho}
\!\frac{\rho_+}{l}\!=\!\frac{2\pi l T}{3}\!\!\left[1+\!\sqrt{1+\!\frac{3}{4(\pi l T)^2}\!\left[1\!+\!(a-1)\!\!\left[\pi l T-\frac{1}{2}\right]^2\right]}\right]\!\!.
\end{equation}
From the above expressions we determine the difference between the free energies of the hRN and hairy black holes 
\begin{equation}\label{delta}
\Delta F= F_{\textrm{hRN}}-F_{\textrm{hairy}}
\end{equation}

In general, it is expected that the free energy is a function of the  two variables of phase space, $T$ and $\Phi$. However, in the hairy black hole,  $\mu$ and $q$ are related (\ref{carga}) and this establishes a relation between $T$ and $\Phi$. For this reason $\Delta F$ in (\ref{delta}) can be expressed in terms of a single variable. In this case, the temperature was chosen. 

We find that the phase diagram has a strong dependence of the parameter $a$, which is proportional to the inverse of the coupling constant $\alpha$. In the Fig. 1, we have illustrated the behavior of  $\Delta F$ for two values of $a$. For all values of $a$, i. e.  $a \geq 1$, there is a phase transition at the critical temperature $T_c= (2\pi l)^{-1}$. This is a second-order phase transition because  $\Delta F$ and its derivative vanish at $T_c$ as it can be seen in Fig. 1, and confirmed by the series expansion of $\Delta F$ around $T_c$
\begin{equation}\label{expansionF}
\Delta F= -\frac{a \sigma \pi l^4 }{64 G}  (T-T_c)^3+\mathcal{O}(T-T_c)^4.
\end{equation}  
In a similar way, as it was discussed for uncharged black holes  \cite{Martinez:2004nb,Koutsoumbas:2006xj}  (the uncharged case corresponds to set $a=1$ here), in the range  $T\geq T_c$ the undressed (hRN) black hole is thermodynamically favored. However, below the critical temperature, the hairy black hole is preferred. 
\begin{figure}
% \centering
\includegraphics[width=8cm]{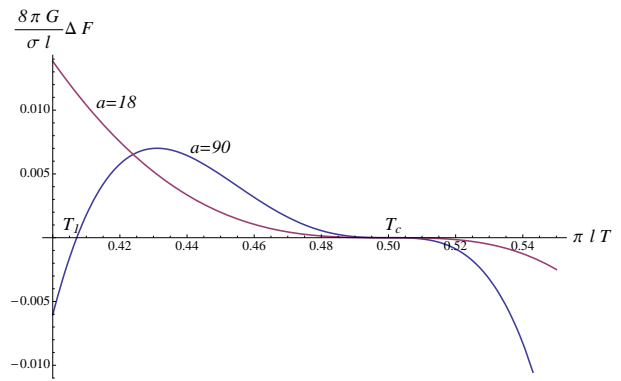}
\caption{ Difference between the free energies, $\Delta F$, of hairy and hRN black holes versus the temperature $T$, computed in the grand canonical ensemble.  The figure consider two cases depending on the values of the parameter $a=2\pi G/(3l^2\alpha)$. In the $a=18$ case, there is a second order-phase transition at $T_c= (2\pi l)^{-1}$. For  the case $a=90$, additionally to the second-order phase transition at $T_c$, there is a first-order one at the temperature $T_1$. }\label{fig1}
\end{figure}

When $ a > a_1= 3(3+2\sqrt3) \sim 19.4$ we find another solution for $\Delta F=0$, denoted by $T_1$,  as it is shown for $a=90$ in Fig. 1 .   This second temperature monotonously increases with $a$ and it asymptotically approaches to $T_c$ (see Fig. 2). However, only when $a>a_2\sim 86$ \footnote{This value, with more precision, is 86.0319986277} this temperature is higher than $T_{min}$, which is the minimum temperature required for ensuring a positive entropy for the hairy black hole. Thus, different to the uncharged case, there is another phase transition provided  $a > a_2$. This new phase transition is of first order because the slope of $\Delta F$ at $T_1$ does not vanish (Fig. 1).   

Additionally, one can includes the extreme hRN black hole into the analysis. For extreme black holes, $\beta$ becomes arbitrary and the entropy vanishes \cite{Teitelboim:1994az}. We set this extreme black hole in the previous ensemble, and using Eq. (\ref{extremo}), the free energy takes the form
\begin{equation}\label{Fext}
F_{\textrm{ext}}=-\frac{\sigma
l}{4\pi G }\left[\frac{1+(a-1)\left(\pi l T-\frac{1}{2}\right)^2}{3}\right]^{3/2}.
\end{equation}
For $a >1$ and $T>0$, $ F_{\textrm{ext}}$ is greater than the free energy corresponding to the non-extremal case, which indicates a thermal decay of the extreme into the non-extreme hRN black hole and none phase transition is observed. This is the same result as in \cite{Martinez:2006an} (see also \cite{Cai:2004pz}). When  $F_{\textrm{ext}}$ and   $F_{\textrm{hairy}}$ are compared, we note that the difference $F_{\textrm{ext}}-F_{\textrm{hairy}}$ vanishes for two different temperatures $T_2$ and $T_3>T_2$, where $T_2$  appears only for $a > a_1$, being $F_{\textrm{ext}}>F_{\textrm{hairy}}$ in the interval $(T_2,T_3)$. This could be a signal of the existence of a phase transition at $T_3$  and another one, for $a >a_1$, at $T_2$. However, since $T_2<T_{min}$ and $T_3 >T_{max}$ (see Fig. 2)
they do not lie within the physical temperature interval $T_{min}<T<T_{max}$.  Therefore, there are no phase transitions between the hairy black hole and the extremal hRN one. Moreover, since $F_{\textrm{ext}}$ is greater than $F_{\textrm{hairy}}$ and $F_{\textrm{hRN}}$  within the proper temperature range, the extremal black hole is able to decay into the hairy or into the non-extremal hRN black hole with different branching ratios.

Finally, as it was mentioned before, the hairy black hole contains an extremal case.  For this extreme black hole, the mass and the charge are fixed in terms of the parameters appearing in the action.  Thus, it is easy to prove that a) only the case of vanishing temperature is compatible with the assumed ensemble, and b) the value of the free energy for the  extreme hairy black hole coincides with the $T=0$ limit of $F_{\textrm{hairy}}$ as it is expected.
%
%\begin{widetext}
\begin{figure}
  \centering
    \includegraphics[width=8.6cm]{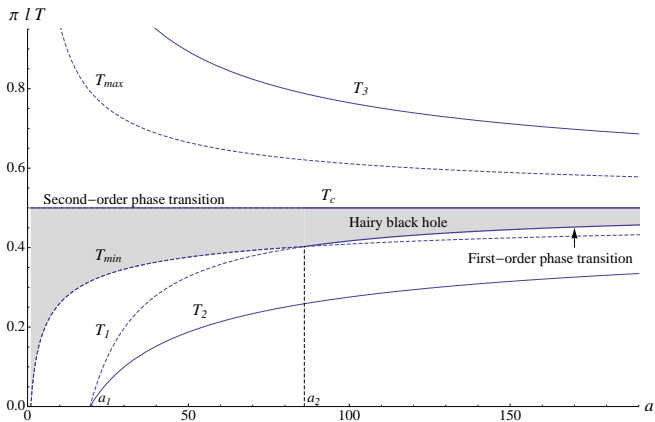}
  \caption{ Phase diagram: The curves for $T_{min}$ and $T_{max}$ enclose the region where the entropy of the hairy black hole is positive. These curves approach to $T_c=(2\pi l)^{-1}$ for $a \rightarrow \infty$. There is  a second-order phase transition at $T_c$ for all values  $a \ge 1$.  For  $a > a_2$ an additional first-order transition appears and it takes place at $T_1$, which is higher than $T_{min}$ within this range of $a$.  The hairy phase (shared region) dominates in the region bounded by $(T_{min}, T_c)$ for $a \le a_2$ , and $(T_{1}, T_c)$ for $a > a_2$. $T_2$ and $T_3$ are out of the physical interval  $(T_{min}, T_{max})$. }\label{fig2}
\end{figure}
%\end{widetext}

\acknowledgments

We thank Andr\'es Gomberoff and Ricardo Troncoso for useful discussions. This work has been partially 
supported by the grants 1085322,  1095098 and 1100755  from Fondecyt and by the Conicyt 
grant ``Southern Theoretical Physics Laboratory" ACT-91 . The Centro de
Estudios Cient\'{\i}ficos (CECS) is funded by the Chilean Government
through the Millennium Science Initiative and the Centers of
Excellence Base Financing Program of Conicyt. CECS is also supported
by a group of private companies which at present includes
Antofagasta Minerals, Arauco, Empresas CMPC, Indura, Naviera
Ultragas and Telef\'onica del Sur. CIN is funded by Conicyt and the
Gobierno Regional de Los R\'{\i}os.

\end{document}